# Examining gender and cultural influences on customer emotions

Dr Vinh Truong, RMIT University, vinh.truongnguyenxuan@rmit.edu.vn

## Abstract

Understanding consumer emotional experiences on e-commerce platforms is essential for businesses striving to enhance customer engagement and personalisation. Recent research has demonstrated that these experiences are more intricate and diverse than previously examined, encompassing a wider range of discrete emotions and spanning multiple-dimensional scales. This study examines how gender and cultural differences shape these complex emotional responses, revealing significant variations between male and female consumers across all sentiment, valence, arousal, and dominance scores. Additionally, clear cultural distinctions emerge, with Western and Eastern consumers displaying markedly different emotional behaviours across the larger spectrum of emotions, including admiration, amusement, approval, caring, curiosity, desire, disappointment, optimism, and pride. Furthermore, the study uncovers a critical interaction between gender and culture in shaping consumer emotions. Notably, gender-based emotional disparities are more pronounced in Western cultures than in Eastern ones, an aspect that has been largely overlooked in previous research. From a theoretical perspective, this study advances the understanding of gender and cultural variations in online consumer behaviour by integrating insights from neuroscience theories and Hofstede's cultural dimension model. Practically, it offers valuable guidance for businesses, equipping them with the tools to more accurately interpret customer feedback, refine sentiment and emotional analysis models, and develop personalised marketing strategies.

## 1. Introduction

E-commerce platforms have revolutionised the way businesses interact with consumers, offering unprecedented opportunities for engagement, data collection, and customer feedback analysis (Yang et al., 2020). These digital marketplaces transcend traditional brick-and-mortar stores, enabling businesses to reach global audiences while facilitating deeper insights into consumer preferences, purchasing behaviour, and satisfaction levels (Zhou et al., 2025). Unlike physical retail environments, where interpersonal communication and nonverbal cues contribute to consumer decision-making, online platforms rely heavily on textual and numerical data to represent customer experiences (Rasappan et al., 2024).

Through written reviews and ratings, customers articulate their experiences, opinions, and emotions, creating a wealth of user-generated content that can be analysed to identify behavioural patterns, emerging trends, and consumer expectations (Karabila et al., 2024). The digital nature of these interactions allows for large-scale data analysis, offering businesses valuable insights into market opinions and product reception. Moreover, by utilising artificial intelligence (AI) and natural language processing (NLP), companies can systematically analyse and interpret customer feedback, including the sentiments and emotions embedded within it (Rasappan et al., 2024).

Sentiments and emotions are essential in shaping consumer perceptions, influencing decision-making, and driving behavioural responses. They determine how individuals assess products, express satisfaction or dissatisfaction, and interact with brands in online environments. At a basic level, analysing customer sentiment helps businesses gauge whether a review reflects a positive or negative opinion about a product or service (Alhadlaq & Alnuaim, 2023). Emotions extend beyond a simple positive-negative binary—they capture not only what customers think but also how they feel, offering deeper insights into their experiences and connections with brands (Guo et al., 2024).

To analyse emotions effectively, researchers have developed various emotion models that categorise and structure emotional states. Traditionally, emotions have been classified into two primary models: categorical and dimensional (Gross, 1998). The categorical model, championed by theorists such as Paul Ekman, posits that humans experience a limited set of basic emotions, including happiness, sadness, anger, fear, surprise, and disgust (Ekman, 1993). These emotions are considered universal across geographical locations, forming the foundation of human affective experiences. In contrast, the dimensional model, as proposed by Russell's circumplex model, suggests that emotions exist along continuous dimensions such as valence (positive to negative) and arousal (low to high) (Posner et al., 2005). This perspective allows for a more nuanced understanding of emotional expression, accommodating variations in intensity and complexity.

Most studies on customer reviews still rely on sentiment analysis or traditional emotion models, either using the six basic categorical emotions or the two-dimensional framework of valence and arousal. However, recent research suggests that emotional experiences are significantly more nuanced. The Hourglass of Emotions model introduces 24 distinct emotional states, offering a more granular perspective on human affect (Susanto et al., 2020). Furthermore, empirical studies have identified even greater complexity, with 27 distinct emotional states emerging from human communication (Cowen & Keltner, 2017; Demszky et al., 2020). Additionally, the conventional two-dimensional model has been found insufficient to capture certain emotional aspects, such as the sense of control or power in an emotional experience. To address these limitations, recent research has proposed a VAD (Valence-Arousal-Dominance) model, which incorporates dominance as a third dimension, allowing for a more comprehensive representation of emotions (Bakker et al., 2014; Xu et al., 2023).

These advancements in theoretical and practical understandings of emotions necessitate a reassessment of customer emotions from multiple perspectives. By moving beyond traditional models and embracing more nuanced frameworks, businesses can gain deeper insights into how emotions shape consumer behaviour (Guo et al., 2024). This reassessment is particularly crucial for tailoring products and services to specific customer demographics, such as gender and cultural background (Kanwal et al., 2022; Mesquita, 2022).

Gender differences in emotional expression have long been explored in psychological and sociological research, yet previous studies have produced conflicting findings (McDonald & Kanske, 2023). Traditionally, societal narratives have portrayed women as more emotionally expressive and men as more reserved, reinforcing gender-based stereotypes in emotional behaviour (Zhao et al., 2018). These stereotypes have shaped expectations regarding how men and women should convey emotions in different settings, including online interactions. However, recent research challenges these conventional perspectives, emphasising that emotional expression is highly context-dependent and shaped by evolving digital communication norms (Pashchenko et al., 2022). Some studies found that while women often report more intense emotional expressions, physiological measures showed no significant gender differences in emotional experience (Deng et al., 2016). With the increasing dominance of digital platforms, there is a growing need to reevaluate gender differences in emotional expression, as individuals now communicate emotions through facial and body language and written language (Kusal et al., 2022).

Similarly, research on cultural differences in online shopping behaviour has produced conflicting findings, particularly concerning Western and Eastern consumers (Hampden-Turner et al., 2020; Wang et al., 2023). Some studies suggest that Western consumers, influenced by individualistic cultural values, tend to provide more detailed, direct, and critical reviews, emphasising personal experiences and product attributes (Marcos-Nájera et al., 2021). In contrast, Eastern consumers, shaped by collectivist cultural norms, may focus on social harmony, indirect feedback, and broader contextual factors such as brand reputation and service quality (Nakayama & Wan, 2018). This cultural divergence reflects fundamental differences in communication styles, value systems, and consumer expectations. However, other studies challenge this binary distinction, noting increasing globalisation and digital standardisation, especially in the e-commerce platforms, which may blur cultural differences in consumer behaviour (Wang et al., 2023).

These conflicting findings highlight the complexity of cultural and gender influences on emotional expression in online reviews, while the interaction between culture and gender remains largely unexplored. Methodological limitations in existing research may contribute to these discrepancies, as many studies rely on sentiment analysis that simplifies emotions into broad categories, often missing subtle distinctions (Acheampong et al., 2020). Others use a restricted set of basic two-dimensional emotions, failing to capture the full spectrum of emotional articulation (Lee et al., 2022), and some analyse reviews from a single platform, product category, or region, limiting the generalizability of their findings (Keller & Kostromitina, 2020). Despite advances in machine learning, three key gaps persist in understanding gender and culture's impact on consumer emotions: research has primarily focused on basic emotions while overlooking more complex categories, studies on dimensional emotions—particularly valence, arousal, and dominance—remain limited, and the interaction and moderating effects of gender and culture in text-based customer reviews are underexplored, leaving a critical aspect of online consumer behaviour unaddressed.

To overcome these limitations, this study utilises a more comprehensive approach to emotional measurement and analysis, incorporating a larger and more diverse dataset. It provides a deeper understanding of how gender and cultural differences influence emotional expression in online reviews. Specifically, this research examines variations in sentiment, valence, arousal, and dominance scores, as well as differences across all sentiment, basic and complex emotion categories. Additionally, it explores the interaction between gender and culture in shaping emotional experiences, offering valuable insights for businesses to refine their targeting strategies with greater precision.

The study aims to address the following three research questions:

1. Do customer emotional experiences differ significantly based on culture?
2. Do customer emotional experiences differ significantly based on gender?
3. Does the impact of gender differences on customer emotional experiences vary across cultures?

By addressing these questions, this study seeks to offer a nuanced understanding of emotional expression in digital communication, drawing on Hofstede's cultural dimensions, neuroscience, and emotion regulation theories to contribute to broader discussions in psychology, cultural studies, and gender research (Gross, 1998; Hines, 2020; Hofstede, 2011). The findings of this study have significant implications for businesses, marketers, and platform developers. By recognising and accommodating cultural and gender-based emotional differences, companies can tailor their communication strategies, optimise user experience, and enhance customer satisfaction.

Additionally, these insights contribute to the development of more inclusive and empathetic digital environments, fostering better interactions between consumers and businesses. Ultimately, this research advances our understanding of the emotional landscape of e-commerce, offering practical applications for improving customer engagement and refining digital communication strategies.

The remainder of this paper is organised as follows: The next section provides a comprehensive literature review on cultural and gender differences in customer emotional experiences, with hypotheses formulated at the end of each subsection. Section 3 outlines the research design and methodology, detailing the data collection and analysis processes. Section 4 presents the study's results and findings, followed by a discussion in Section 5, which explores the significance of the findings, study limitations, and directions for future research. Finally, Section 6 concludes the paper.

## 2. Literature Review

This section examines existing literature on cultural and gender differences in customer emotional experiences, highlighting key findings and conflicting perspectives. It explores how cultural norms influence emotional expression in online reviews, particularly in sentiment, valence, arousal, and dominance scores, as well as across basic and advanced emotion categories. Additionally, it investigates gender-based variations in emotional articulation, considering both traditional stereotypes and evolving digital communication norms. By identifying gaps in current research—such as inconsistencies in prior findings, methodological limitations, and the need for multidimensional emotion analysis—this review lays the foundation for formulating hypotheses that will be empirically tested in this study.

## 2.1. Cultural Differences

Businesses operate across diverse cultural contexts, making it essential to understand how culture influences customer emotional experiences. While some studies suggest that cultural norms significantly shape emotional expression, others present conflicting evidence, highlighting the complexity of this relationship. These inconsistencies underscore the need for further empirical research to clarify the role of culture in shaping consumer emotions.

One of the most influential frameworks in cross-cultural emotion research is Hofstede's cultural dimension theory, which differentiates cultures based on dimensions such as individualism-collectivism and uncertainty avoidance (Hofstede, 2011). According to this theory, individualistic cultures, such as those in Western countries, encourage emotional expressiveness and personal opinions, while collectivistic cultures, prevalent in Eastern societies, emphasise emotional restraint and social harmony (D. Matsumoto et al., 2008; Vonk & Silva, 2024). This distinction suggests that customers from Western cultures may express stronger sentiments, whether positive or negative, while those from Eastern cultures may moderate their emotional expressions to maintain group cohesion (Hampden-Turner et al., 2020).

Some empirical studies seem to confirm these theoretical predictions. For instance, a large-scale analysis of online product reviews found that American consumers tend to use more emotionally intense language in their feedback compared to Japanese consumers, whose reviews were more neutral and restrained (Nakayama & Wan, 2018). Similarly, another study analysing Weibo and Twitter posts revealed that Western social media users exhibited greater emotional variability, while Eastern users displayed more moderate and context-dependent emotional expressions (Filieri & Mariani, 2021).

However, conflicting evidence exists. Some researchers argue that globalisation and digital communication are gradually reducing cultural differences in emotional expression (Yang et al., 2020). A recent study comparing Chinese and American customer reviews on multinational e-commerce platforms found that while some traditional cultural patterns persisted, the differences in sentiment and emotion scores were significantly smaller than expected (Long & Lei, 2022). Similarly, Klaus Scherer and colleagues conducted a study across 37 countries, revealing that the structure and intensity of emotional experiences were strikingly similar across cultures, reinforcing the notion that core emotional processes are universal (Scherer et al., 2001).

Furthermore, the role of dataset quality and diversity in shaping emotional expression cannot be ignored. Studies have shown that English allows for more explicit emotional expression, whereas languages like Chinese and Japanese rely more on contextual cues and implicit emotional communication (Alhadlaq & Alnuaim, 2023). Similarly, studies have found that on platforms where reviews are encouraged to be highly detailed (e.g., Amazon), cultural differences in emotional expression are more pronounced, whereas on platforms with character limits (e.g., Twitter), differences tend to diminish (Wu, 2021). Research has shown that cultural differences in emotional expression are more prominent in hedonic product reviews (e.g., entertainment, fashion) than in utilitarian product reviews (e.g., electronics, household goods) (Cortis, 2021).

The challenge to the conventional understanding of cultural differences in emotional expression comes from studies that consider the role of emotional granularity. Emotional granularity refers to an individual's ability to differentiate between distinct emotional states (Barrett & Lida, 2024). Some researchers have suggested that cultural differences in emotional expression are less about intensity and more about the specificity of emotions conveyed. For instance, Eastern consumers may be less likely to express high-arousal emotions like excitement or anger but may instead articulate more complex emotions such as nostalgia or gratitude (Wang et al., 2020). This perspective implies that analysing basic emotions alone may not capture the full extent of cultural differences in customer experiences.

Furthermore, recent advances in emotion detection using artificial intelligence (AI) and natural language processing (NLP) have enabled researchers to classify emotions beyond the traditional valence-arousal models. Studies applying these techniques have revealed that Western consumers exhibit higher dominance scores in their emotional expressions, aligning with cultural tendencies that emphasise personal agency and self-expression. In contrast, Eastern consumers often exhibit lower dominance scores, reflecting cultural values of humility and interdependence (Yik et al., 2023).

Some scholars argue that methodological biases in data collection and emotion classification may contribute to inconsistent findings. For example, studies relying solely on sentiment analysis may oversimplify emotional expression by categorising emotions as simply positive or negative, failing to capture subtle cultural variations (Kusal et al., 2022). Others contend that studies using predefined emotion categories may impose Western-centric emotional frameworks that do not fully align with the emotional lexicon of Eastern cultures (Yang et al., 2020). Some studies have begun using EEG to capture more data on emotions accurately, but they face challenges with scalability (Li et al., 2024).

To address these limitations, some studies have adopted a more comprehensive approach by integrating sentiment scores with advanced emotion categories, such as admiration, optimism, disappointment, and pride. A cross-cultural study examining social network communications found that Western consumers were more likely to express emotions such as excitement and pride, while Eastern consumers showed higher tendencies for emotions like gratitude and disappointment (Cordaro et al., 2018). This suggests that cultural differences extend beyond simple positive-negative sentiment classifications and require a multidimensional analysis.

Given these theoretical and empirical insights, this study proposes the following hypothesis:

**H1: Customer emotional experiences differ significantly based on culture, as measured by sentiment, valence, arousal, and dominance scores, as well as sentiment, basic, and advanced emotion categories.**

## 2.2. Gender Differences

Emotional experiences play a crucial role in consumer behaviour, influencing purchasing decisions, brand perception, and overall engagement. One of the most debated questions in psychological and marketing research is whether emotional expression and experience differ significantly between genders.

Neuroscience studies have explored sex differences in brain structure and function, highlighting the impact of factors such as brain size and hormonal influences. Research also identifies sex-based variations in the neurobiology of social behaviour, memory, emotions, and brain injury recovery, emphasising the crucial role of estrogens in regulating forebrain function (Choleris et al., 2018).

Accordingly, traditionally, women have been perceived as more emotionally expressive, while men are often characterised as more emotionally restrained (Fischer et al., 2018; Ward & King, 2018). However, recent empirical studies present mixed findings, suggesting that gender differences in emotional experiences may not be as clear-cut as previously assumed (Li et al., 2024).

Thelwall (2018) found that women express more positive emotions in online reviews, reinforcing the stereotype that they are more emotionally expressive. However, other research suggests that sentiment differences are minimal in digital settings, as online anonymity reduces gendered communication norms (Pashchenko et al., 2022).

Similarly, Tsai and Clobert (2019) found that women were more likely to describe experiences in emotionally charged terms, whereas men's responses remained more balanced. However, other research contradicts this pattern, showing that valence differences are less pronounced in text-based online interactions (Derks et al., 2011).

In one of the studies, women have been found to experience both positive and negative emotions more intensely than men (Lausen & Schacht, 2018). This aligns with broader psychological literature suggesting that women exhibit higher physiological and subjective emotional responses. However, Kret and De Gelder (2012) argued that men also experience high arousal in competitive or achievement-oriented contexts, such as when reviewing technology or sports-related products (Truong, 2023).

For dominance-related emotions, traditional studies suggest that men exhibit higher emotions, such as confidence and pride, while women display more submissive emotions, such as sadness or anxiety (Grossman & Wood, 1993). However, newer studies indicate that online environments enable women to express more assertive emotions, reducing historical gender disparities in dominance (Löffler & Greitemeyer, 2023).

For basic emotions, some research suggests that women experience and express sadness and fear more frequently, whereas men are more likely to express anger (Chaplin & Aldao, 2013). However, in online communication, gender differences in basic emotions appear less pronounced (Bhattacharya et al., 2021). Given the shift from face-to-face to digital interactions, it is unclear whether traditional gender differences in basic emotions persist in consumer reviews.

Apparently, despite the extensive research on gender differences in emotional expression, several limitations remain in existing studies which caused the conflicts. First, many rely on sentiment analysis, which often oversimplifies emotional experiences by categorising them as positive, negative, or neutral. This approach overlooks the complexity of emotional expression, especially in online consumer reviews, where emotions are often nuanced and context-dependent (Nandwani & Verma, 2021). While some studies suggest that women express more positive emotions and men adopt a more critical tone, others argue that sentiment differences diminish in digital communication due to platform-specific norms and varying levels of anonymity (Ligthart et al., 2021). This inconsistency raises concerns about the reliability of sentiment-based gender comparisons.

Second, while valence and arousal provide more granular insights into emotional experiences, studies on these dimensions yield conflicting results. Some research suggests that women express emotions with higher valence extremes and greater intensity, while men exhibit more neutral or controlled expressions (Yik et al., 2023). Furthermore, dominance, traditionally linked to assertiveness in men and submissiveness in women, appears to be evolving in digital spaces, where women increasingly display assertive emotions (Xu et al., 2023). These inconsistencies suggest that current models may not fully capture the fluidity of emotional expression across different online platforms and consumer segments.

Third, studies examining basic emotions such as happiness, anger, and sadness have shown gender differences in face-to-face interactions, yet these differences appear less pronounced in digital settings. While traditional research suggests that women express more sadness and fear and men more anger, newer studies indicate that online communication blurs these distinctions (Alvarez-Gonzalez et al., 2021). The shift from verbal and nonverbal cues in face-to-face interactions to text-based digital communication challenges conventional emotional expression theories. This highlights the need for research that considers how gender differences manifest in written language rather than relying on traditional psychological models developed for offline interactions.

Given the conflicting evidence from previous studies, a multidimensional approach is necessary to comprehensively examine gender differences in emotional experiences. Measuring sentiment, valence, arousal, dominance, and both basic and advanced emotion categories allows for a more robust understanding of gendered emotional expression. This approach moves beyond simplistic sentiment analysis, providing a clearer picture of how men and women express emotions in online consumer reviews.

Based on the literature review and theoretical considerations, the following hypothesis is proposed:

**H2: Customer emotional experiences differ significantly based on gender, as reflected in sentiment, valence, arousal, and dominance scores, as well as basic and advanced emotion categories.**

## 2.3. Gender and cultural differences

The intersection of gender and culture in emotional responses has been a central theme in cross-cultural psychology and gender theories. Theories such as Display Rules Theory (Ekman &

Friesen, 1969) and Emotion Regulation Theory (Gross, 1998) provide frameworks for understanding how cultural norms and gender expectations shape emotional expression and experience. Social Role Theory (Eagly, 1987) also posits that societal expectations for men and women are based on traditional roles—men as assertive and stoic and women as nurturing and emotionally expressive. The Emotional Communication Framework (LaFrance & Banaji, 1992) highlights that women are generally more attuned to both expressing and interpreting emotions, a trend observed across cultures but varying in degree.

These theories emphasise the significance of the interaction between gender and culture, offering explanations for observed differences in certain facial emotions and non-verbal communications. The difference was primarily observed along the individualism-collectivism dimension.

For example, while both men and women in collectivistic cultures (e.g., Japan, China) may suppress outward displays of negative emotions to maintain social harmony, women might still exhibit more subtle emotional cues than men due to gendered expectations of nurturing and empathy (David Matsumoto et al., 2008). In contrast, individualistic cultures (e.g., the United States and Canada) promote open emotional expression, potentially reducing gender differences in emotional displays.

Similarly, research by (Fischer et al., 2004) found that in collectivistic societies, women reported higher levels of guilt and shame, emotions that promote social cohesion, while men reported emotions linked to status and dominance, such as pride and anger. Conversely, in individualistic cultures, these gender differences may be less distinct, as emotional expression is more individualised and less constrained by traditional roles.

In individualistic societies, such as the U.S. or Western Europe, where self-expression is encouraged, men and women alike may exhibit a broader range of emotional displays. However, women still tend to be more emotionally expressive, particularly in non-verbal communication, such as facial expressions and body language (Brody, 2008). In collectivistic societies, like those in East Asia, emotional expression is more subdued, but women may still display emotions more freely within close social circles, maintaining gendered patterns of emotional communication (David Matsumoto et al., 2008).

Safdar et al. (2009) conducted a cross-cultural study involving participants from 23 countries and found that while women consistently reported greater emotional expressiveness than men, the magnitude of this difference varied by culture. In individualistic cultures, such as Canada and the U.S., gender differences were smaller, while in collectivistic cultures, like Japan and South Korea, the differences were more pronounced.

However, recent studies have questioned the interaction between culture and gender, suggesting that globalisation and technological advancements have reduced differences between men and women across countries.

For example, some studies found that gender differences in emotional experiences are stronger in collectivistic cultures, not in previous ones (Brody & Hall, 2010). Additionally, studies using sentiment analysis and dimensional emotion models have found mixed results regarding whether gender differences in emotional expression are consistent across cultures or whether they diminish in digital communication settings where traditional social norms are less influential (Kuppens & Verduyn, 2017).

Some other studies could not find any difference in emotions due to the interaction between culture and gender and claimed that it is due to the weak effect of one of the two factors. For example, a study analysed self-reported emotional experiences and expressions among men and women in different regions in the U.S. to common stereotypes, the findings revealed no significant gender differences in the frequency of everyday subjective feelings, suggesting that men and women experience emotions similarly regardless of their locations (Simon & Nath, 2004).

Another research examined gender differences in emotional expression across various contexts and developmental stages. The study found that while some differences exist, they are often small and context-dependent, indicating that both men and women can express a wide range of emotions, and these expressions are influenced more by situational factors than by gender alone (Chaplin, 2015).

These conflicting results can be explained by some of the limitations in previous research. One is the reliance on simplistic sentiment analysis or basic emotion models that categorise emotions into binary positive or negative states. Such methods fail to capture the complexity of emotional experiences, which include variations in valence, arousal, and dominance (Bakker et al., 2014). Moreover, many studies focus on a limited set of emotions, overlooking the broader spectrum of basic and advanced emotional states, such as admiration, disappointment, and pride (Cowen & Keltner, 2017). Additionally, most prior research has examined gender and cultural differences separately without thoroughly investigating their interaction. This gap highlights the need for a more nuanced analysis that considers both sentiment and multidimensional emotional experiences across gender and cultural groups.

Despite growing interest in understanding emotional expression in online consumer behaviour, limited research has explored how gender and culture interact to shape emotional experiences in e-commerce settings. Existing studies have primarily examined gender or cultural differences in isolation, neglecting their potential combined influence (Wang et al., 2023). Furthermore, while recent advances in natural language processing and emotion detection have enabled more precise measurement of emotional experiences, few studies have applied these techniques to investigate the interplay between gender and culture in online customer reviews (Hossain et al., 2022).

Based on the theoretical and empirical gaps identified, this study proposes the following hypothesis:

**H3: The impact of gender differences on customer emotional experiences varies across cultures, as measured by sentiment, valence, arousal, and dominance scores, as well as sentiment, basic, and advanced emotion categories.**

By examining these hypotheses, this study seeks to provide a more comprehensive understanding of cultural and gender-based emotional expression in digital communication and consumer behaviour. The next section will introduce the conceptual model underpinning this research, outlining the hypothesised direct and moderating effects, the data collection and analysis methods, and the rationale for selecting these techniques to evaluate the proposed hypotheses.

## 3. Methodology

This section will introduce the conceptual model for the study, followed by a discussion on the data collection process and the statistical techniques employed for analysis.

### 3.1. Conceptual Model

Based on the hypotheses formulated in the previous literature section, the conceptual model presented in Figure 1 illustrates the intricate relationships between gender, culture, and emotional experiences in consumer behavior. It posits three key hypotheses: H1, suggesting that gender influences emotional experiences; H2, indicating that culture has an impact on emotional experiences; and H3, hypothesizing that gender and culture interact to produce distinct emotional responses. This framework provides a comprehensive lens to explore how demographic factors shape consumer emotions, particularly in the context of customer reviews, offering valuable insights into personalized marketing and customer engagement strategies.

The first hypothesis (H1) posits that cultural background plays a pivotal role in shaping emotional experiences. Culture influences not only the type of emotions people experience but also how they express and interpret them. For instance, Eastern cultures are believed to emphasize collective values and emotional restraint, leading to more subdued expressions of emotions, while Western cultures tend to encourage individualism and emotional expressiveness (Hofstede, 2011). By examining sentiment scores, dimensional emotions like valence, arousal, and dominance, and categorical emotions such as basic and complex emotions, this model allows for a detailed analysis of cultural variances in emotional expression within customer reviews.

The first hypothesis (H2) asserts that gender significantly influences emotional experiences. This is rooted in extensive psychological and marketing literature in neuroscience, which has shown that men and women often differ in how they perceive, express, and respond to emotional stimuli. For instance, women are believed to express emotions more openly and are more sensitive to emotional cues, while men may exhibit more restrained emotional responses (Hines, 2020). In this model, emotional experiences are further dissected into three categories: Sentiment, Dimensional Emotions, and Categorical Emotions, allowing for a nuanced examination of how gender affects each dimension of emotional response in consumer reviews.

The third hypothesis (H3) highlights the interaction between gender and culture in influencing emotional responses. This suggests that the emotional experiences of male and female consumers are not uniform across cultures; rather, they are shaped by the cultural norms and expectations associated with gender roles. For instance, while Western women may feel more comfortable expressing dissatisfaction openly, Eastern women might express the same sentiment in a more indirect manner according to the Emotion Regulation Theory (Gross et al., 2006). Similarly, cultural expectations may influence how men from different regions display emotions like frustration or satisfaction. This interaction is critical for understanding the nuanced dynamics of emotional responses in diverse consumer populations.

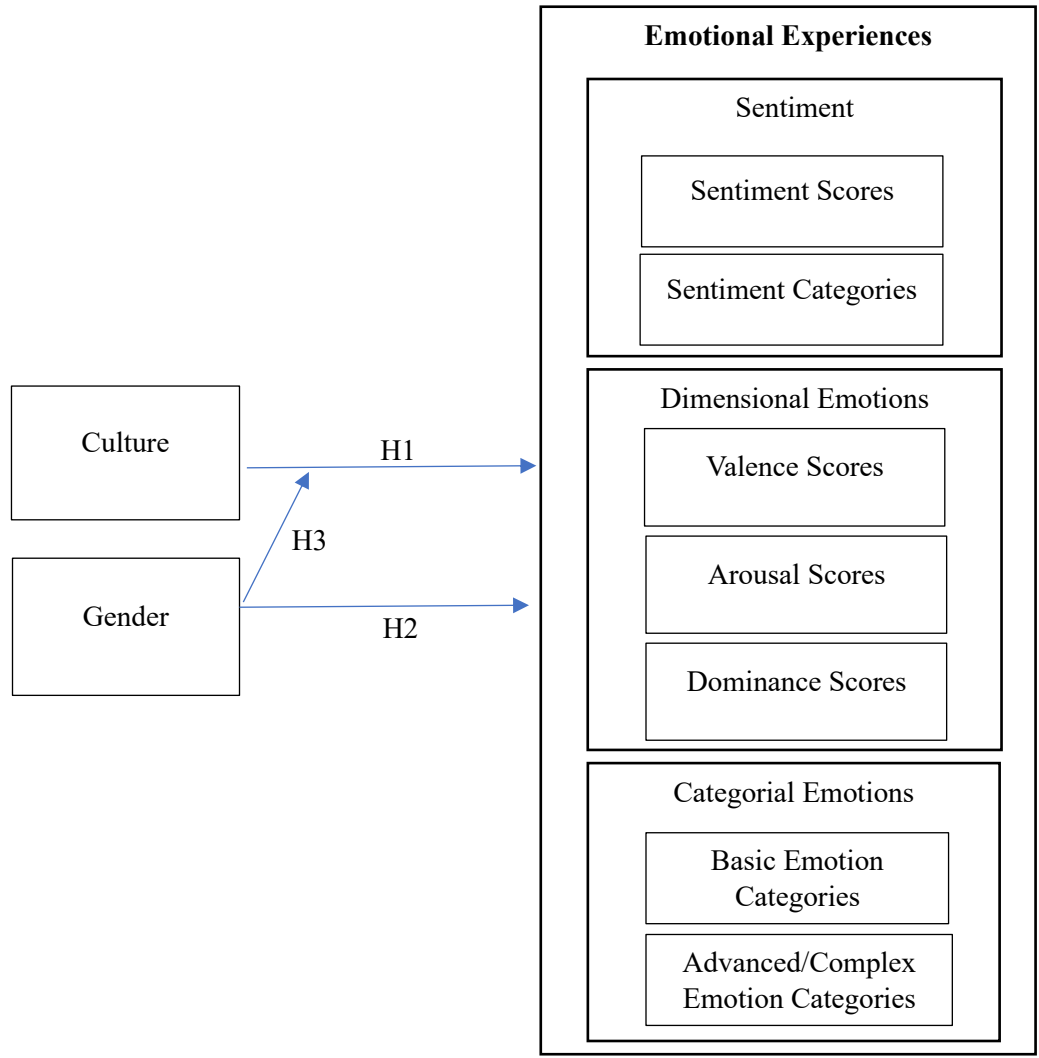

*Figure 1: The conceptual model of this study*

The model categorizes emotional experiences into three main areas: Sentiment, Dimensional Emotions, and Categorical Emotions. Sentiment analysis involves sentiment scores and sentiment categories (positive, negative, neutral), providing a broad measure of overall emotional tone (Hutto & Gilbert, 2014). Dimensional emotions, based on models like Russell's Circumplex Model of Affect (Russell, 2003), include valence (pleasure-displeasure), arousal (intensity of emotion), and dominance (control over the emotion) (Bakker et al., 2014). Lastly, categorical emotions classify responses into basic emotions (e.g., happiness, anger, fear) and complex emotions (e.g., pride, guilt) (Demszky et al., 2020). This multi-layered approach allows for a comprehensive understanding of the depth and complexity of consumer emotional responses.

By integrating gender and culture into the analysis of emotional experiences, this model offers valuable insights for consumer behavior research. This conceptual model sets the stage for exploring the complex interplay of gender and culture in shaping emotional experiences. By examining customer reviews through this lens, researchers can uncover patterns that might otherwise remain hidden in more generalized analyses. Incorporating both sentiment analysis and nuanced emotional dimensions provides a comprehensive examination of the emotional landscape in consumer feedback, helping to address recurring conflicts found in previous studies.

## 3.2. Data collection

In pursuit of the mentioned objectives, this study proposed a comprehensive method for extracting and categorizing customer emotions, consisting of several steps. The workflow of the suggested method is visualized in Figure 2.

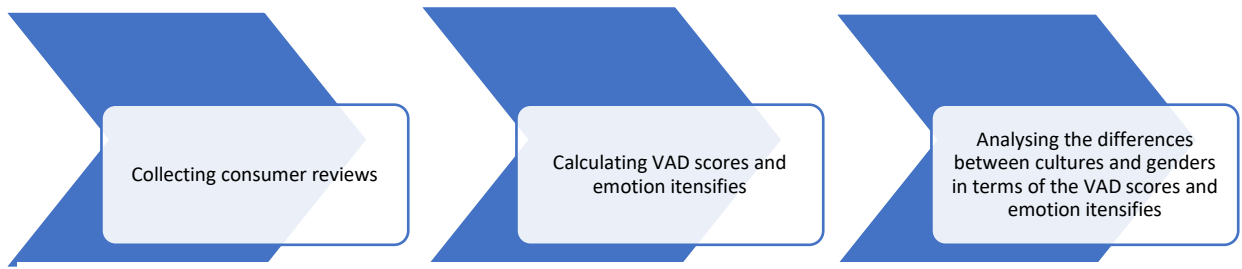

*Figure 2: Proposed method workflow*

For the initial stage of gathering customer experiences, this study focused on collecting reviews from open-source platforms. The primary requirement for these reviews was the inclusion of both gender and cultural background information of the reviewers. This specific criterion is crucial to accurately analyze the interaction between gender and cultural influences on emotional responses, which is a key focus of the research. By targeting datasets that contain these demographic details, the study aims to ensure that the subsequent analysis is both comprehensive and reflective of diverse consumer perspectives across different platforms, industries and products.

Despite the abundance of customer reviews available online across various platforms, one of the significant challenges encountered was the lack of comprehensive demographic information in many of these reviews. While some datasets included gender, they often lacked cultural or nationality details, and vice versa. This limitation required an extensive search and careful filtering of potential data sources to ensure that only those reviews meeting the study's criteria were selected. This process underscored the difficulties in finding datasets that are both rich in content and detailed in demographic attributes, highlighting a gap in the availability of comprehensive consumer feedback data.

These datasets span various platforms, industries and products, providing a broad perspective on consumer behavior across different contexts. The datasets include FashionNova (fashion), Amazon Reviews (electronics), Celsius Network (cryptocurrency trading), ASOS TrustPilot (cosmetics), Qatar Airways (airline services), and La Veranda (restaurant and hotel). The FashionNova Customer Review Dataset comprises a collection of customer feedback and ratings gathered from the popular online fashion retailer, FashionNova. This dataset includes detailed textual reviews, star ratings, timestamps, and customer demographic information such as gender, age, and location when available. The reviews cover a wide range of products, from clothing and accessories to footwear, providing insights into customer satisfaction, product quality, fit, and overall shopping experience.

*Table 1: Datasets with gender and culture information*

| No | Dataset | Category | Number of raw records | Number of clean records |
|---|---|---|---|---|
| 1 | Fashionnova | Fashion | 131980 | 90884 |
| 2 | Amazon Reviews | Electronics | 21214 | 19728 |
| 3 | Celsius Network | Cryptocurrency trading | 21966 | 12297 |
| 4 | ASOS TrustPilot | Cosmetics | 4020 | 3295 |
| 5 | Qatar Airlines | Airline | 2369 | 2364 |
| 6 | La Veranda | Tourism, Hospitality | 1528 | 729 |

Among social networks, Amazon stands out as a platform that can offer user-generated reviews and comments linked to gender-disclosed profiles (Haque et al., 2018). Additionally, Amazon provides robust datasets that are accessible for research purposes, making it a valuable resource for analyzing text-based communication patterns. The Amazon customer review data encompasses millions of reviews from users worldwide, offering insights into how consumers interact with products and services (AlQahtani, 2021). Each review typically includes a star rating, written feedback, and metadata such as the date of the review and the reviewer's profile information (Haque et al., 2018).

The Celsius Network Customer Review Dataset consists of user-generated feedback and ratings from customers using the Celsius Network platform, a cryptocurrency lending and borrowing service. This dataset includes textual reviews, star ratings, and occasionally user demographics, offering insights into customer satisfaction regarding platform usability, interest rates, security, and customer support (AlQahtani, 2021). Similarly, the ASOS TrustPilot Customer Review Dataset contains reviews from a pilot group of ASOS TrustPilot customers, focusing on their experiences with specific product lines, delivery services, and the overall shopping interface. The dataset captures detailed product feedback, fit, style preferences, and customer service interactions, along with demographic information like age, gender, and location when available (Parker & Alexander, 2022).

The Qatar Airways Customer Review Dataset comprises detailed feedback from passengers regarding their experiences with the airline, covering aspects such as in-flight services, comfort, customer support, punctuality, and overall satisfaction. Reviews include star ratings, textual comments, and traveler demographics like nationality, age, and travel class, providing a rich resource for analyzing customer sentiment and service quality in the aviation industry (Hamad MA Fetais et al., 2021). Similarly, the La Veranda Customer Review Dataset features reviews from patrons of La Veranda, a well-known restaurant, capturing feedback on food quality, ambiance, service, and pricing. This dataset includes star ratings, textual descriptions, and occasionally customer details such as visit frequency and group size y (Kondopoulos, 2014).

The customer reviews apparently randomly sampled from various dataset categories. This diversity in categories helps capture a wide range of contexts and communication styles, reducing the risk of skewed results caused by overrepresentation from specific domains (Hewson et al., 2016). By ensuring that both variables are present in the data, the study can perform a nuanced analysis that captures these differences directly and interactively in subsequent sections.

Once the datasets were collected, a thorough data cleaning process was undertaken to ensure the integrity and usability of the data. This process involved the removal of unnecessary fields, such as metadata unrelated to the review content, and the correction of broken characters and formatting issues often present in raw textual data. Additionally, duplicate entries and incomplete records lacking essential demographic information were eliminated. After the data cleaning process, a total of 129,297 customer reviews are ready for sentiment and emotion detection.

### 3.3. Sentiment and Emotion Detection

Once the customer reviews are collected, advanced machine learning tools are next to be used to analyze them, generating sentiment, valence, arousal, and dominance scores. Additionally, the reviews will undergo a detailed examination to identify both basic and complex emotions, along with their intensities.

As shown in Table 2, various machine learning models are employed to analyse customer reviews across multiple dimensions of emotional expression. The first task, converting text into sentiment scores, is performed using tabularisai/multilingual-sentiment-analysis, a model designed to classify sentiment intensity across different languages. This model assigns a numerical sentiment score to each review, helping determine the degree of positivity, negativity, or neutrality in consumer feedback. By utilising a multilingual approach, the model ensures robust sentiment evaluation across diverse linguistic and cultural backgrounds, making it particularly useful for cross-cultural studies in customer emotion analysis that not many sentimental analysis model can do (Rasappan et al., 2024).

*Table 2: Machine learning models*

| No | Task | Model name |
| --- | --- | --- |
| 1 | Text to the sentiment scores | tabularisai/multilingual-sentiment-analysis |
| 2 | Text to sentiment categories | vaderSentiment |
| 3 | Text to valence, arousal and dominance scores | hplisiecki/word2affect_english |
| 4 | Text to basic emotion categories | bhadresh-savani/bert-base-uncased-emotion |
| 5 | Text to advanced emotion categories | SamLowe/roberta-base-go_emotions |

For the second task, which involves classifying text into sentiment categories, the popular VADER (Valence Aware Dictionary and sEntiment Reasoner) model is used. VADER is a rule-based model specifically designed for sentiment analysis in social media, product reviews, and other text-based data sources (Hutto & Gilbert, 2014). Unlike traditional sentiment analysis tools, VADER takes into account the contextual intensities of words, as well as punctuation, capitalization, and emoticons, making it highly effective for analyzing informal and opinion-rich texts. Its ability to accurately detect sentiment in short and conversational texts makes it ideal for processing customer reviews, where emotional expressions may be subtle yet impactful.

To measure valence, arousal, and dominance (VAD) scores, the study utilizes hplisiecki/word2affect_english, a model that is recently developed to assign numerical values representing the emotional tone (valence), intensity (arousal), and level of control (dominance) in a given text. These three dimensions provide a deeper understanding of emotional expression beyond simple sentiment classification (Plisiecki & Sobieszek, 2024). Valence reflects the positivity or negativity of an emotion, arousal measures the level of emotional excitement, and dominance assesses the degree of control or power associated with the emotion. By leveraging this advanced model, the study can capture the nuanced emotional experiences of consumers and examine their differences in 3-D scales.

The fourth task involves categorizing text into basic emotions, which include happiness, sadness, anger, fear, surprise, and disgust. To accomplish this, the study employs bhadresh-savani/bert-base-uncased-emotion, a transformer-based model fine-tuned for emotion classification. This model is trained on extensive datasets to accurately assign basic emotion labels to text, allowing researchers to explore fundamental emotional expressions in consumer reviews (Khalili et al., 2022). By classifying emotions at this level, the study can analyze whether certain consumer groups, such as men or women, express specific basic emotions more frequently and whether these patterns vary across cultures.

Finally, to classify text into advanced emotion categories, the study uses SamLowe/roberta-base-go_emotions, a model fine-tuned on the GoEmotions dataset (Demszky et al., 2020). This dataset includes 27 distinct emotion labels, such as admiration, amusement, pride, disappointment, and gratitude, providing a highly granular view of emotional expression. Advanced emotion classification allows for a more sophisticated analysis of consumer sentiment, revealing complex emotional states that may not be captured by basic emotion models. By incorporating this model, the study aims to uncover intricate patterns in how different demographics express nuanced emotions in digital communication, enhancing the understanding of consumer behavior in a multicultural and gender-diverse landscape, something that previous studies has not successfully performed.

### 3.4. Statistical techniques

Once the machine learning model identified and categorized the emotions, statistical techniques were applied to examine how these emotions varied across different cultural and gender groups. Descriptive statistics, including means and standard deviations, were first used to assess data distribution and ensure normality. The dataset contains a relatively equal representation of male and female consumers, as well as Western and Eastern reviews, allowing for balanced comparisons. Additionally, key emotional metrics—sentiment score, valence, arousal, and dominance scores—were evaluated for skewness and kurtosis, confirming that their distributions fell within acceptable limits.

The use of a large dataset, comprising 129,297 customer reviews, strengthened the reliability of the analysis by minimizing biases and ensuring a more representative sample. The normality of the data was further supported by the even distribution of sentiment categories, as well as basic and complex emotion classifications. These preliminary statistical checks provided a solid foundation for further inferential analyses, enabling more accurate comparisons of emotional expression across different demographic and cultural segments.

To further explore differences in emotional expression across cultural and gender groups, inferential statistical techniques were employed. Univariate tests were conducted to determine whether significant differences existed in the distribution of emotions between these groups. To gain deeper insights into the complex relationships between culture, gender, and emotional intensity, multivariate analysis techniques were applied. Multivariate Analysis of Variance (MANOVA) was used to simultaneously examine multiple dependent variables, such as sentiment scores and emotion categories, and their variations across cultural and gender groups. This allowed for a more comprehensive assessment of how these factors interact (Saunders, 2015). Additionally, the moderation regression test using Hayes' PROCESS macro was employed to investigate whether culture moderated the effect of gender on emotional expression (Hayes, 2017). By incorporating moderation analysis, the study is able to determine whether cultural differences amplified, diminished, or altered gender-based emotional patterns, offering a more nuanced perspective on how emotions are shaped by both demographic factors.

The findings from these inferential analyses are presented in the following section, highlighting key statistical results and their implications.

## 4. Analysis Results

This section presents the findings from the multivariate and moderation regression analyses, organized based on the results of the hypothesis tests. Additionally, any findings beyond the scope of the hypotheses will be discussed further.

### 4.1. Hypothesis 1 Testing

The first hypothesis assumes that there is a strong relationship between culture and customer emotional experiences. Table 3 outlines the results of hypothesis testing aimed at examining the effect of Culture on various emotional and sentiment-related variables. The data highlights significant cultural influences on several dependent variables, with notable differences in how culture affects specific emotional dimensions. For example, Culture has a strong, statistically significant impact on Sentiment Scores, with a Type III Sum of Squares of 28.527, an F-value of 52.380, and a significance level (p-value) of .000. This result suggests that cultural background plays a crucial role in shaping the overall sentiment intensity individuals express, pointing to cultural norms or communication styles that influence how sentiments are conveyed.

The Sentiment Category, which classifies sentiments into five categories including very positive, positive, neutral, negative and very negative, also shows a statistically significant effect from cultural differences, albeit to a lesser degree. The F-value of 3.400 with a p-value of .033 indicates that culture does affect how people categorize their sentiments, though this influence is less pronounced compared to the sentiment intensity. This could imply that while cultures may share similar frameworks for categorizing emotions, the nuances of how these categories are applied vary subtly across cultural contexts.

*Table 3: Hypothesis 1 testing result*

| Independent Variable | Dependent Variable | Type III Sum of Squares | df | Mean Square | F | Sig. |
|---|---|---|---|---|---|---|
| Culture | Sentiment Score | 28.527 | 2 | 14.264 | 52.380 | .000 (***) |
| | Sentiment Category | 12.481 | 2 | 6.241 | 3.400 | .033 (*) |
| | Valence Scores | 4.205 | 2 | 2.103 | 66.493 | .000 (*) |
| | Arousal Score | .355 | 2 | .178 | 43.566 | .000 (*) |
| | Dominance Score | 2.815 | 2 | 1.408 | 136.873 | .000 (*) |
| | Basic Emotion Category | .417 | 2 | .208 | .193 | .825 |
| | Advanced Emotion Categories | 4844.616 | 2 | 2422.308 | 37.772 | .000 (*) |

A particularly striking finding is the significant influence of culture on Valence, Arousal, and Dominance scores, which represent core emotional dimensions. Valence Scores, reflecting the positivity or negativity of an emotion, have an F-value of 66.493 and a p-value of .000, demonstrating that cultural context heavily shapes the emotional tone individuals experience or express. Similarly, Arousal Scores (F = 43.566, p = .000) show a strong cultural effect, suggesting that the level of emotional excitement or calmness varies considerably across different cultures. The most significant result is seen in Dominance Scores, with an F-value of 136.873 and a p-value of .000, indicating that cultural background profoundly affects feelings of control or submission in emotional experiences.

Interestingly, the analysis reveals that Basic Emotion Categories are not significantly influenced by culture. The F-value of 0.193 and the p-value of .825 suggest that basic emotions such as anger, joy, or sadness are relatively universal and consistent across cultures. This aligns with the idea that basic emotions are biologically hardwired and less susceptible to cultural modulation. These findings imply that while the intensity and expression of emotions might differ between cultures, the fundamental experience of basic emotions remains consistent.

In contrast, Advanced Emotion Categories demonstrate a significant cultural influence, with a Type III Sum of Squares of 4844.616, an F-value of 37.772, and a p-value of .000. Advanced emotions, such as admiration, remorse, or pride, are more context-dependent and influenced by social norms, values, and interpersonal dynamics, which vary greatly across cultures. This significant effect suggests that complex emotional experiences are deeply intertwined with cultural background, likely reflecting differing cultural interpretations, emotional regulation strategies, and value systems. The inclusion of complex emotion categories has revealed the impact of culture on customer emotional experiences, an effect that was not detected using basic emotion categories in previous studies.

The gender differences in customer emotional experiences are particularly evident when analyzing the complex emotion categories, as illustrated in Figure 3.

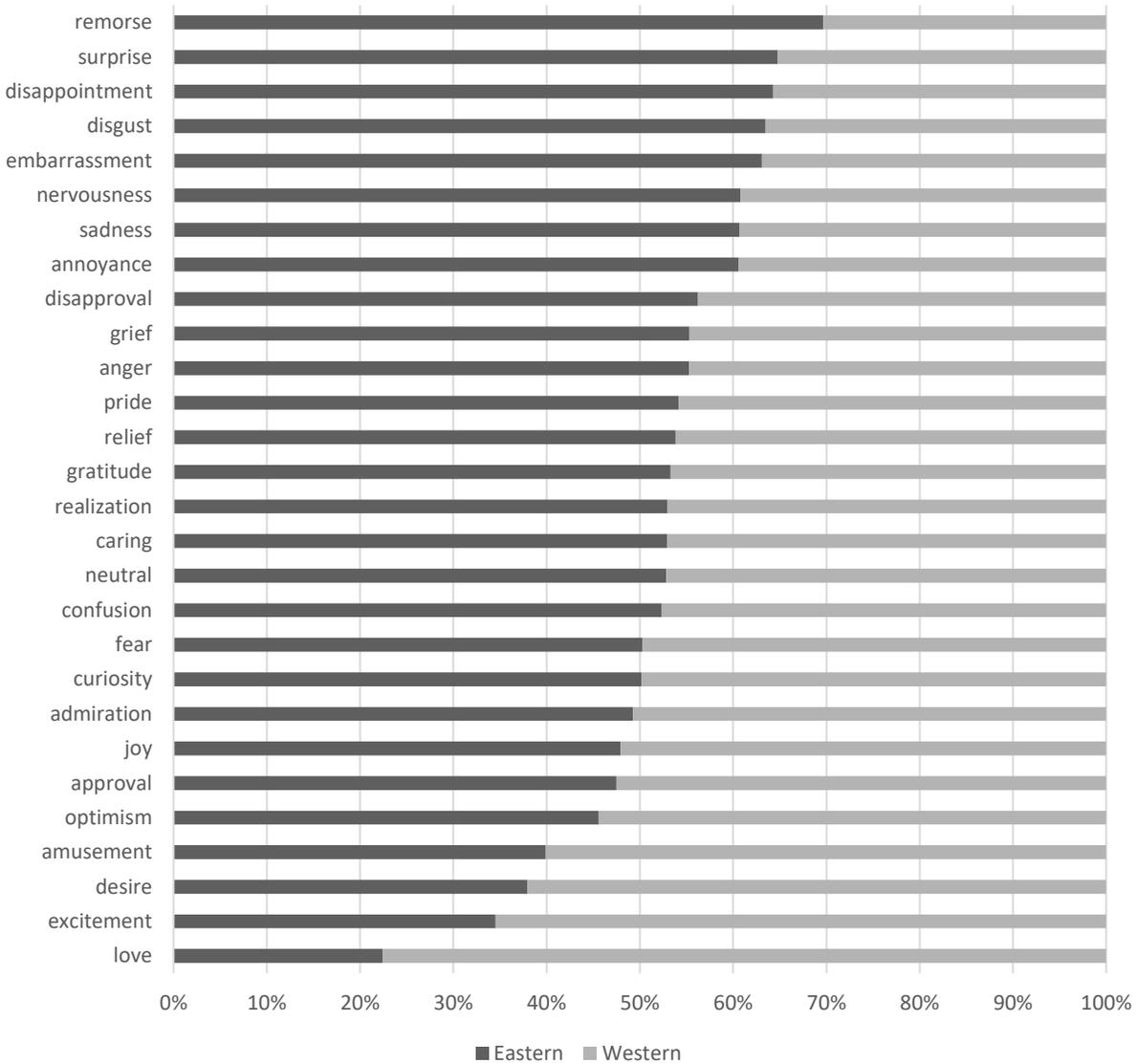

*Figure 3: Cultural differences in advanced emotion categories*

In Figure 3, both groups exhibit high levels of admiration and neutral responses, with Western participants slightly more likely to report admiration (22.4% vs. 21.7%), while Eastern participants lean more towards neutral responses (21.4% vs. 19.1%). Notably, approval is more prevalent in Western responses (8.7% compared to 7.8%), suggesting a possible cultural inclination towards positive reinforcement in the West. Conversely, emotions like annoyance and disappointment are significantly higher among Eastern participants (7.2% and 10% respectively) than their Western counterparts (4.7% and 5.6%), indicating potential differences in how dissatisfaction is experienced or expressed.

Positive emotions such as love and excitement display stark contrasts between the two cultures. Western participants report a much higher incidence of love (14.7% vs. 4.2%) and excitement (1.3% vs. 0.7%), suggesting that Western cultures may be more openly expressive of strong positive emotions. Similarly, desire is more common in the Western group (1.5% vs. 0.9%), which could reflect cultural differences in openness regarding personal aspirations and wants. On the other hand, Eastern participants show higher levels of negative emotions such as disgust (1.3% vs. 0.8%), embarrassment (0.4% vs. 0.3%), and sadness (1.5% vs. 1.0%), indicating a greater prevalence or willingness to report these emotions within Eastern cultural contexts.

Interestingly, emotions like curiosity, fear, and joy are relatively consistent across both groups, with only slight variations. For example, curiosity and fear are nearly identical in both cultures, while joy is slightly higher among Western participants (2.8% vs. 2.6%). Additionally, both groups exhibit low levels of grief, pride, and nervousness, suggesting these emotions may not be as prominently triggered in the contexts examined. The data highlights how cultural factors influence emotional expression and perception, with Eastern cultures potentially leaning towards more complex or mixed emotional states, while Western cultures might favor more overt expressions of strong positive emotions like love and excitement.

Overall, the statistical analysis has detected a strong relationship between culture and emotional expression. While basic emotions appear to transcend cultural boundaries, more complex emotions and sentiment dimensions are significantly shaped by cultural context. These findings underscore the importance of considering cultural factors in psychological research and emotional analysis, especially in customer reviews, as they can profoundly influence how emotions are experienced, expressed, and categorized.

### 4.2. Hypothesis 2 Testing

The results of the Hypothesis 2 test differ slightly from those of the Hypothesis 1 test. As shown in Table 4, the analysis reveals that gender does not have a significant impact on Sentiment Scores, with a Type III Sum of Squares of 0.204, an F-value of 0.749, and a p-value of 0.387. This suggests that the overall intensity of sentiment expressed by individuals does not vary significantly between genders. Similarly, Sentiment Category—which groups sentiments into broad categories like positive, negative, or neutral—also shows no significant gender differences, indicated by an F-value of 2.008 and a p-value of 0.156. These results imply that while individual expressions of emotion might differ, the overall sentiment and its categorization are relatively consistent across genders.

dimensions of Valence, Arousal, and Dominance. Valence Scores, which reflect the positivity or negativity of an emotion, show a strong gender effect, with a Type III Sum of Squares of 0.878, an F-value of 27.781, and a highly significant p-value of 0.000. This indicates that men and women may experience or express emotions with differing levels of positivity or negativity. Similarly, Arousal Scores, which measure the level of emotional excitement or calmness, are significantly affected by gender (F = 24.743, p = 0.000). This suggests that men and women may differ in how intensely they experience emotions, with potential implications for understanding gender-specific emotional responses.

*Table 4: Hypothesis 2 testing result*

| Independent Variable | Dependent Variable | Type III Sum of Squares | df | Mean Square | F | Sig. |
|---|---|---|---|---|---|---|
| Gender | Sentiment Score | .204 | 1 | .204 | .749 | .387 |
| | Sentiment Category | 3.686 | 1 | 3.686 | 2.008 | .156 |
| | Valence Scores | .878 | 1 | .878 | 27.781 | .000 (***) |
| | Arousal Score | .101 | 1 | .101 | 24.743 | .000 (***) |
| | Dominance Score | .263 | 1 | .263 | 25.542 | .000 (***) |
| | Basic Emotion Category | 3.222 | 1 | 3.222 | 2.977 | .084 |
| | Advanced Emotion Categories | .005 | 1 | .005 | .000 | .993 |

However, the influence of gender becomes significant when examining the emotional Dominance Scores also exhibit a significant gender difference, with an F-value of 25.542 and a p-value of 0.000. Dominance in emotional contexts refers to feelings of control or submission, and this result indicates that gender plays a role in how individuals perceive their emotional agency. These findings align with existing research suggesting that societal norms and expectations can influence how different genders express emotions related to power and control. The strong significance across valence, arousal, and dominance suggests that while general sentiment may not differ by gender, the qualitative aspects of emotional experiences do.

In contrast, the effect of gender on Basic Emotion Categories is not statistically significant, though it approaches the conventional threshold (F = 2.977, p = 0.084). Basic emotions, such as anger, joy, sadness, or fear, appear to be experienced similarly across genders, suggesting that these fundamental emotional responses are less influenced by gender differences. This supports the idea that basic emotions are biologically rooted and universal, although slight variations may exist in expression or frequency.

Interestingly, Advanced Emotion Categories show no significant gender effect at all, with an F-value of 0.000 and a p-value of 0.993. Advanced emotions, which include more complex and socially constructed feelings like admiration, remorse, or pride, seem to be equally distributed across genders in this sample. This lack of difference could suggest that, despite cultural or societal expectations about emotional expression, men and women experience complex emotions in comparable ways.

Statistical data in Figure 4 shows that certain emotions are expressed more frequently by males than females.

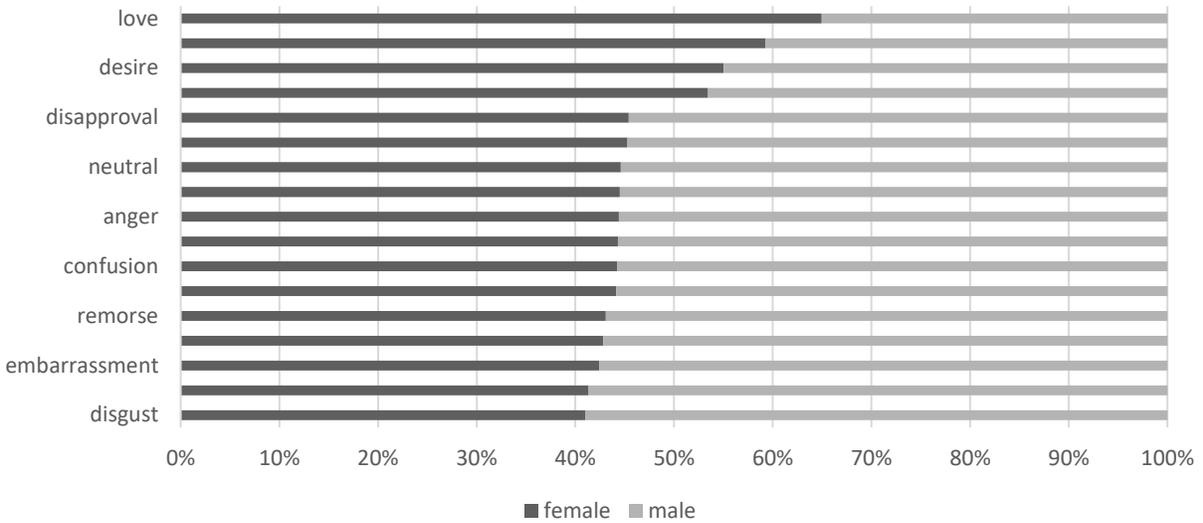

*Figure 4: Gender differences in advanced emotion categories*

In Figure 4, females exhibit slightly higher levels of admiration (22.8% compared to 21.9%), love (17.3% vs. 9.3%), and neutral responses (17.6% vs. 21.8%). On the other hand, males show stronger expressions of anger (5.8% compared to 4.1%), disgust (4.4% vs. 3.7%), and disappointment (6.5% vs. 5.2%).

A deeper dive into the data shows that males tend to report slightly higher levels of negative emotions such as annoyance (9.0% vs. 8.5%), disapproval (1.0% vs. 0.7%), and sadness (1.1% vs. 0.9%). Conversely, females exhibit stronger positive emotions like joy (2.9% vs. 2.6%) and gratitude (4.1% vs. 3.9%). Interestingly, curiosity and desire are reported equally at 0.5% and slightly higher for females in the case of desire (1.6% vs. 1.3%).

The most significant gender disparity is observed in the love and neutral categories. Females report nearly double the love responses compared to males (17.3% vs. 9.3%), while males show a considerably higher percentage of neutral responses (21.8% vs. 17.6%). This could imply that females are more emotionally expressive or responsive in contexts that evoke affection, whereas males may maintain a more neutral stance in similar situations.

In summary, the statistical analysis indicates that while gender does not significantly affect overall sentiment intensity or basic emotional categories, it plays a notable role in shaping specific emotional dimensions such as valence, arousal, and dominance. These results highlight the complexity of gender differences in emotional expression, suggesting that while the fundamental experience of emotions may be similar, the nuances of emotional tone, intensity, and control differ between men and women. This underscores the importance of considering gender as a moderating factor in emotional research, particularly when examining the qualitative aspects of emotional experiences.

## 4.3. Hypothesis 3 Testing

Hypothesis 3 aimed to examine the interaction between Gender and Culture variables and identify the moderating factor in this relationship. Graphs will be plotted to visually illustrate the moderating effect of one factor on another.

### 4..3.1. Multivariate Analysis

Table 5 presents the results of the hypothesis testing that examines the interaction effect of Gender and Culture on various emotional and sentiment-related variables. The interaction between gender and culture significantly influences Sentiment Scores, with a Type III Sum of Squares of 18.283, an F-value of 33.570, and a p-value of 0.000. This suggests that the combined effect of cultural background and gender plays a crucial role in determining the overall intensity of sentiments expressed. While gender alone did not show a significant impact on sentiment in the previous analysis, this result indicates that cultural factors, when combined with gender, create variations in how sentiments are experienced and conveyed.

*Table 5: Gender and Cultural interaction*

| Independent Variables | Dependent Variable | Type III Sum of Squares | df | Mean Square | F | Sig. |
|---|---|---|---|---|---|---|
| Gender * Culture | Sentiment Score | 18.283 | 2 | 9.141 | 33.570 | .000 (***) |
| | Sentiment Category | .765 | 2 | .383 | .208 | .812 |
| | Valence Scores | 1.100 | 2 | .550 | 17.398 | .000 (***) |
| | Arousal Score | .015 | 2 | .008 | 1.844 | .158 |
| | Dominance Score | .241 | 2 | .120 | 11.713 | .000 (***) |
| | Basic Emotion Category | 7.103 | 2 | 3.552 | 3.282 | .178 |
| | Advanced Emotion Categories | 410.030 | 2 | 205.015 | 3.197 | .041 (*) |

However, the interaction effect of gender and culture on Sentiment Category is not statistically significant, as indicated by an F-value of 0.208 and a p-value of 0.812. This suggests that the categorization of sentiments into broad groups like positive, negative, or neutral is not substantially affected by the interplay of gender and cultural factors. This result implies that while the intensity of sentiment might vary with gender and culture, the basic way sentiments are classified remains consistent across different demographic groups.

The interaction effect is particularly significant for Valence Scores, with a Type III Sum of Squares of 1.100, an F-value of 17.398, and a p-value of 0.000. This indicates that gender and culture together influence the positivity or negativity of emotions experienced by individuals. This interaction suggests that cultural norms and gender roles may shape emotional valence differently in various cultural contexts. For instance, certain cultures may encourage more positive emotional expression in one gender over the other, leading to observable differences in valence scores when both factors are considered together.

On the other hand, the interaction between gender and culture does not significantly affect Arousal Scores, with an F-value of 1.844 and a p-value of 0.158. Arousal, which measures the intensity or excitement level of emotions, appears to be relatively unaffected by the combined influence of gender and culture. This suggests that, while arousal levels might differ individually due to gender or cultural factors, their interaction does not produce significant variations. This could imply a more universal response to emotionally arousing situations that transcends cultural and gender differences.

The interaction effect is significant for Dominance Scores, with an F-value of 11.713 and a p-value of 0.000. This indicates that the combined influence of gender and culture plays an important role in how individuals perceive feelings of control or submission in emotional contexts. Cultural norms regarding power dynamics and gender roles likely contribute to this variation. For example, cultures with more hierarchical gender roles might exhibit stronger gender-based differences in dominance perceptions, while more egalitarian cultures might show fewer differences.

Finally, the interaction effect is also significant for Advanced Emotion Categories. The F-value for advanced emotions, it is 3.197 ($p = 0.041$). This suggests that gender and cultural factors interact to influence the expression of complex emotions. For emotions like joy, anger, or sadness, cultural expectations about gender-appropriate emotional expression may shape how these feelings are outwardly displayed. Similarly, emotions like admiration, approval and caring, which are more nuanced and socially constructed, are also influenced by the intersection of gender and culture, reflecting the intricate ways that societal norms and individual identities intertwine to shape emotional experiences.

The Multivariate Analysis identified four interactive effects, partially confirming the interaction between culture and gender in shaping customer emotional experiences. This finding is the necessary condition to confirm the moderating effects of one factor on another with the next regression analysis.

### 4.3.2. Moderation Regression Analysis

The moderation regression analysis utilizes Hayes' PROCESS macro to continue assessing the moderating effects of Gender and Culture on their relationship with Customer Emotional Experiences. Visual representations will be used to illustrate these moderating effects more clearly.

*Moderating the effect on Sentiment Score*

The moderation regression test using Hayes' PROCESS macro results presented in its model summary indicate a statistically significant, albeit modest, overall model fit. The R value of 0.0888 and R-squared of 0.0079 suggest that the model explains approximately 0.79% of the variance in the dependent variable. The F-statistic of 342.1925 ($p < .0001$) demonstrates that the model is statistically significant, meaning that the predictors—Gender, Culture, and their interaction (Gender x Culture)—collectively contribute to predicting the outcome variable. The test of the highest-order interaction reveals an R² change of 0.0004 with an F-value of 51.1246 (p

< .0001), indicating that even though the additional variance explained by the interaction term is small, it is still statistically significant. This highlights the moderating role of gender in influencing the relationship between culture and the dependent variable.

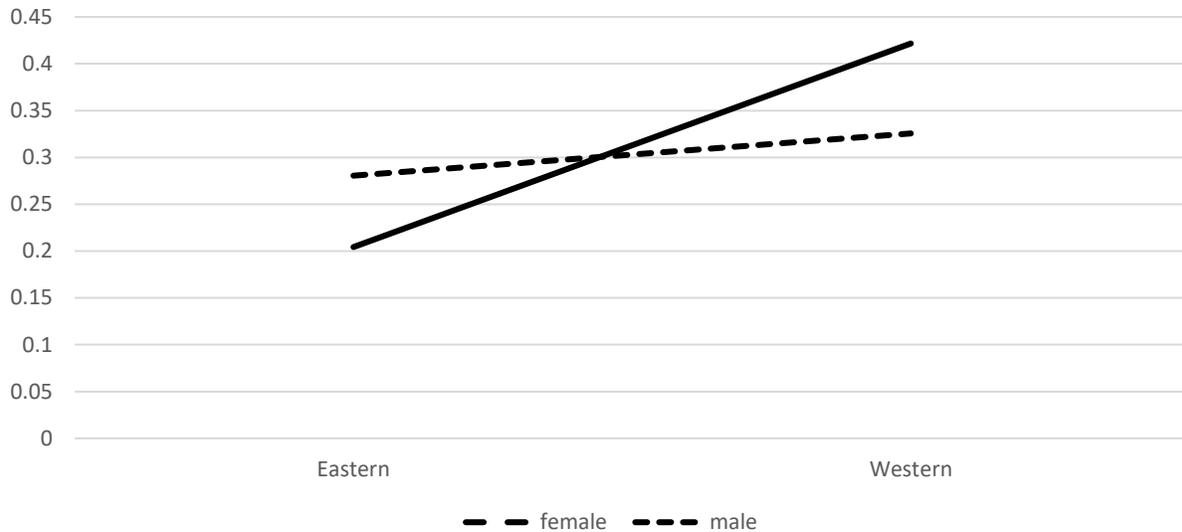

*Figure 5: Moderating effects on the sentiment score*

As shown in Figure 5, in the Eastern category, male sentiment scores are higher than female scores. Males have a sentiment score close to 0.28, while females have a lower score of approximately 0.21. This suggests that, within Eastern contexts, males exhibit more positive sentiments than females. The gap between the two groups in this category is the largest among the three, indicating a significant gender-based variation in sentiment.

In the Western category, the trend shifts, with female sentiment scores surpassing those of males. Females reach a sentiment score of about 0.42, while males drop to around 0.33. This reversal highlights that in Western contexts, females exhibit more positive sentiments than males, contrasting with the trend observed in Eastern contexts. This shift could reflect cultural differences in how emotions are expressed or perceived between genders in different parts of the world. Gender clearly moderated the relationship between culture and sentiment score.

*Moderating the effect on Valence score*

The moderation regression analysis using Hayes' PROCESS macro for the outcome variable Valence reveals a statistically significant model, with an R value of 0.1099 and an R-squared of 0.0121. This indicates that the model explains approximately 1.21% of the variance in Valence, which, while modest, is statistically significant ($F = 526.9640$, $p < .0001$). The test of the highest-order interaction confirms the importance of this interaction, with an $R^2$ change of 0.0002 and an F value of 22.7202 ($p < .0001$). The conditional effects analysis further illustrates this interaction by showing that at high levels of Culture (Culture = 3.0000), the effect of Gender on Valence becomes significantly negative ($\beta = -0.0401$, $p < .0001$). This suggests that gender moderates the effect of culture on valence scores.

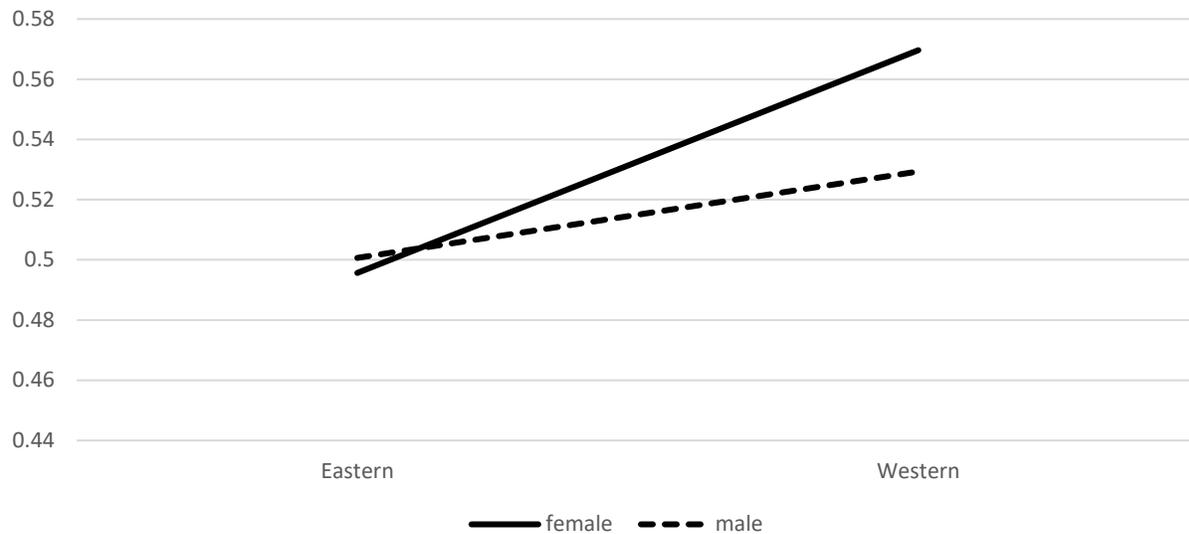

*Figure 6: Moderating effects on the valence score*

The graph in Figure 6 illustrates the emotional valence—or the positivity of emotional responses—of male and female groups across two cultural categories: Eastern, and Western. The vertical axis ranges from 0.44 to 0.58, indicating subtle variations in emotional positivity. The straight line represents female valence scores, while the dotted line represents male scores. This visualization reveals differences in how emotional valence shifts between genders in different cultural contexts.

In the Eastern category, both male and female valence scores are closely aligned, hovering around the 0.50 mark. Females show a slightly lower score than males, indicating a marginally less positive emotional valence. The small gap between genders in this category suggests that emotional expressions in Eastern contexts are relatively consistent across males and females, reflecting possible cultural norms that emphasize similar emotional regulation for both genders.

In the Western category, female valence scores continue to rise slightly, peaking near 0.57, while male scores decrease to around 0.53. This indicates that females in Western contexts tend to express more positive emotions compared to their male counterparts. The divergence between the two genders is most pronounced in this category, highlighting cultural influences that may encourage or allow for greater emotional expressiveness in females compared to males in Western societies.

## Moderating the effect on the dominance score

This moderation regression analysis using Hayes' PROCESS macro investigates the relationship between Gender (the predictor) and Dominance (the outcome variable), moderated by Culture (the moderator). The model accounts for a small but statistically significant portion of the variance in Domin ($R^2 = 0.0114$, $F(3, 129293) = 494.96$, $p < .001$). The conditional effects analysis demonstrates that at higher levels of Culture (e.g., Culture = 3.0), the relationship between Gender and Dominance becomes negative and statistically significant ($b = -0.0203$, $p <$

.001). This suggests that in cultural contexts with higher Culture values, increases in Gender are associated with decreases in Dominance. Gender clearly moderated the effect of culture on the dominance score.

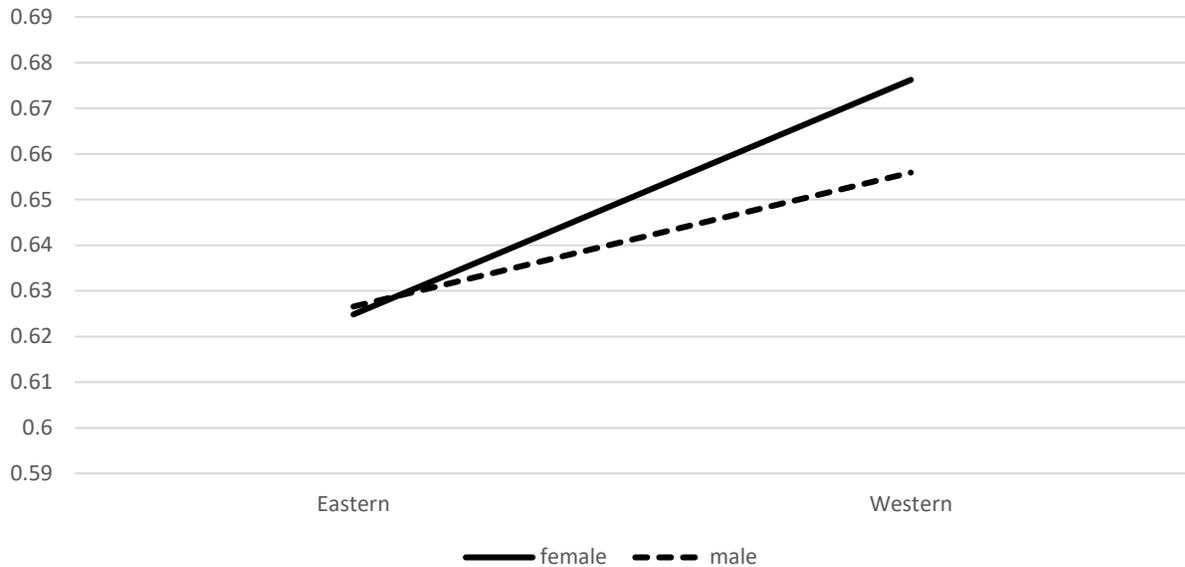

*Figure 7: Moderating effect on the dominance score*

The graph in Figure 7 illustrates how male and female participants perceive or experience dominance across Eastern and Western cultural contexts. Dominance scores typically reflect feelings of control, influence, or authority in a given context, with higher scores indicating stronger perceptions of dominance. Both the female (straight line) and male (dotted line) groups exhibit an upward trend in dominance scores from Eastern to Western contexts, but females consistently score slightly higher than males as the cultural context shifts.

In the Eastern context, both male and female dominance scores are nearly identical, hovering around 0.63. This close alignment suggests that in Eastern cultural settings, there may be a more balanced perception of dominance between genders. The relatively modest scores could also reflect cultural values that emphasise collectivism or hierarchical structures, where dominance might be less individually asserted or equally distributed across genders.

In the Western context, the trend continues, with dominance scores rising slightly for both genders. Females peak at around 0.675, while males reach about 0.657. The widening gap between female and male dominance scores suggests that Western cultural settings might foster greater perceptions of control and influence among women. This could be attributed to social norms in Western societies that emphasise individualism, equality, and personal agency, particularly in promoting gender parity. Overall, the graph highlights how cultural context plays a significant role in shaping dominance perceptions, with a consistent edge for females as we move from Eastern to Western environments.

## Moderating Effects on Advanced Emotion Categories

This moderation regression analysis using Hayes' PROCESS macro examines the relationship between Gender (the predictor) and Advanced Emotion Categories, with Culture serving as the moderator. The overall model is statistically significant ($R^2 = 0.0011$, $F(3, 129293) = 46.33$, $p < .001$), but the proportion of variance explained is minimal, indicating that the model, while statistically significant, has limited practical explanatory power. The conditional effects analysis reveals that at higher levels of Culture (Culture = 3.0), the effect of Gender on Advanced Emotion Categories becomes significantly negative ($b = -0.4067$, $p < .001$). This indicates that in stronger cultural contexts, increases in generational characteristics are associated with a decrease in competence characteristics. Gender can moderate the effect of culture on advanced emotion categories.

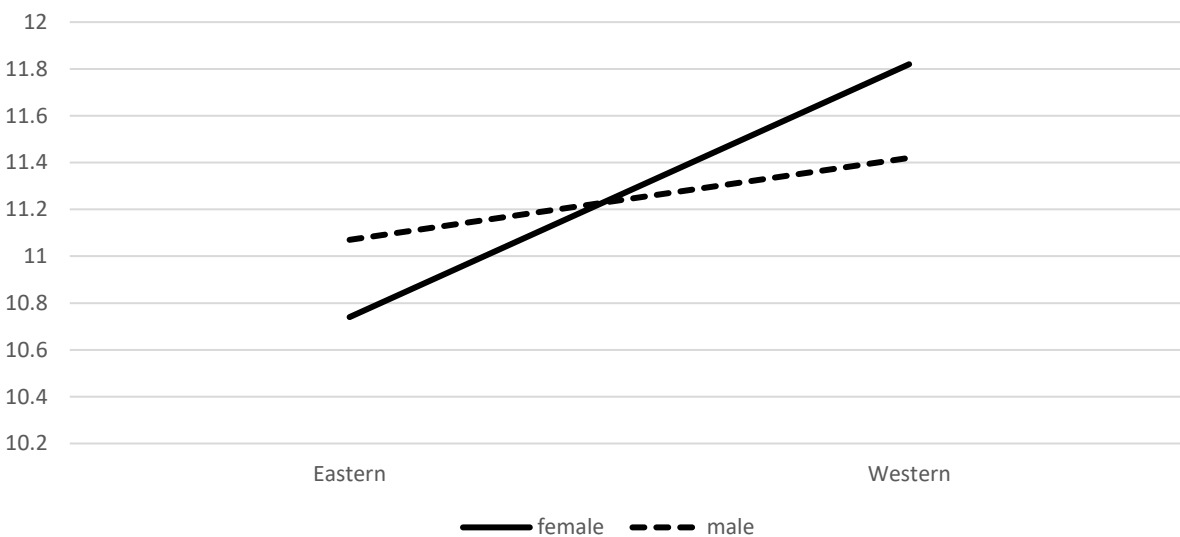

*Figure 8: Moderating the effect on Advanced Emotion Categories*

Figure 8 illustrates how male and female participants score across two cultural contexts: Eastern and Western. Complex emotions, which include feelings like pride, guilt, embarrassment, and empathy, are often influenced by cultural norms and social interactions. Initially, in the Eastern context, males display slightly higher scores (around 11.1) compared to females (approximately 10.7). This suggests that men might experience or express complex emotions more prominently in Eastern settings, possibly due to cultural factors influencing emotional roles and expectations.

In the Western context, the scores for both genders rise significantly, with females reaching approximately 11.8 and males around 11.4. The sharp increase, particularly for females, indicates that Western cultural norms may encourage a more open acknowledgement and expression of complex emotions. The fact that female scores surpass male scores in this context suggests that Western societies might foster greater emotional awareness and expression among women, aligning with broader cultural values that emphasise individuality and emotional authenticity.

Overall, the analysis highlights the dynamic interplay between gender and cultural context in shaping complex emotional experiences. While males exhibit higher scores in the Eastern context, the shift towards Western culture reverses this trend, with females ultimately displaying a greater intensity of complex emotions. This pattern underscores how deeply cultural frameworks can affect emotional dynamics across genders.

## 5. Discussion
### 5.1. Key Findings

As summarised in Table 6, this section presented varying levels of support for the proposed hypotheses, with some being fully confirmed, others partially supported, and a few not showing significant effects. Hypothesis 1, which examined the impact of Culture on emotional experiences, is fully confirmed, as Culture significantly influences sentiment scores, sentiment categories, valence, arousal, dominance, and advanced emotion categories. This finding suggests that cultural background plays a crucial role in shaping emotional expression in consumer reviews.

*Table 6: Hypothesis testing results*

| Hypothesis | Independent variable | Sentiment Score | Sentiment Category | Valence Scores | Arousal Score | Dominance Score | Basic Emotion CCategory | Advanced Emotion Category | Remark |
|---|---|---|---|---|---|---|---|---|---|
| 1 | Culture | Yes | Yes | Yes | Yes | Yes | | Yes | Western is higher than Eastern. |
| 2 | Gender | | | Yes | Yes | Yes | | | Females are higher than males. |
| 3 | Culture x Gender | Yes | | Yes | | Yes | | Yes | The difference between Western and Eastern is even higher for females than male |

Hypothesis 2, which investigated the effect of Gender, is partially confirmed. While Gender significantly impacts valence, arousal, and dominance scores, it does not show a strong influence on sentiment scores, sentiment categories, or basic or advanced emotional expressions. This partial confirmation indicates that gender-based differences in emotional expression exist but may not be as pronounced in certain dimensions, potentially due to the influence of digital communication norms.

Hypothesis 3, which tested the interaction between Culture and Gender, shows mixed results. The interaction effect is confirmed for sentiment scores, valence, dominance and advanced emotion categories, indicating that cultural differences in emotional expression are moderated by gender. However, no significant interaction effects are observed for sentiment categories, arousal, or basic emotion categories. Gender did moderate the effects of culture on customer emotional experiences, and in most cases, the effects are significantly higher for Western customers compared with their Eastern counterparts.

Several findings in this study diverge from those reported in previous literature—for example, the absence of significant gender differences in basic emotion categories and the inconsistent patterns observed in sentiment polarity. One possible explanation lies in the nature of the data itself: written, text-based online reviews. Unlike spoken or multimodal communication, textual data may limit the richness of emotional expression due to linguistic constraints, character limits, or the tendency toward neutral or formal language in consumer-generated content (Kusal et al., 2022). As a result, certain affective signals may be muted, and the full intensity or subtlety of emotion may not be adequately conveyed.

In addition, cross-cultural and gendered variations in emotional expression are often embedded in metaphor, idiom, and culturally specific phrasing. These nuances are difficult for automated natural language processing models to detect, especially when relying on simple sentiment scores or basic emotion labels. Conventional emotion detection tools—often limited to positive, negative, and neutral classifications or Ekman's six basic emotions—may overlook the more complex affective profiles found in real-world data. This is particularly problematic in cross-cultural contexts, where emotional norms and expressive styles vary considerably.

To address these limitations, this study adopted a more nuanced approach by employing Valence-Arousal-Dominance (VAD) scoring and a 27-category emotion model (based on GoEmotions), rather than relying solely on sentiment polarity or a limited set of basic emotions. This allowed for a richer and more differentiated analysis of emotional content, capturing subtle differences that are more aligned with the multifaceted nature of human affect. Nonetheless, given the inherent constraints of the data and tools, some theoretical implications drawn from the findings remain interpretive. They are grounded in established literature and theory, while acknowledging that the complexity of emotional communication—particularly across cultures and genders—may not be fully captured in text alone.

## 5.2. Theoretical contributions

The study's findings contribute significantly to Hofstede's cultural dimensions theory by reinforcing the notion that cultural differences shape emotional experiences. The results demonstrate that Western individuals exhibit higher sentiment scores, valence, arousal, and dominance compared to their Eastern counterparts, aligning with Hofstede's framework (Hofstede, 2011). Specifically, these findings support the distinction between individualistic and collectivistic cultures, where Western individualistic societies encourage open emotional expression, while Eastern collectivistic cultures promote emotional restraint (Vonk & Silva, 2024). This provides empirical support for the role of cultural values in influencing emotional responses and highlights the necessity of considering cultural context in studies of sentiment and emotion.

From a neuroscience perspective, the study advances the understanding of how gender differences influence brain mechanisms underlying emotional processing. The observed variations in sentiment and emotional intensity suggest potential neurobiological differences in affective regulation across genders. For example, research in affective neuroscience has shown that gender differences impact the neural pathways involved in emotion perception and expression, particularly in brain regions such as the amygdala and prefrontal cortex (Hines, 2020). The findings align with previous studies suggesting that males, who emphasise individual expression, may exhibit stronger neural activation in areas associated with high-arousal emotions, whereas females may engage regulatory mechanisms that suppress overt emotional responses (Barrett & Lida, 2024).

The study also contributes to emotion regulation theory by providing empirical evidence that cultural and gender factors influence how emotions are processed and expressed. Emotion regulation theory posits that individuals use various cognitive and behavioural strategies to modulate their emotional experiences. The findings suggest that Eastern individuals may engage in greater emotional suppression, whereas Western individuals may favour expressive strategies, supporting Gross (1998) model of emotion regulation. Additionally, gender differences in emotional responses align with research indicating that women tend to use more emotion-focused regulation strategies, such as cognitive reappraisal, compared to men, who may rely more on suppression or distraction (Butler et al., 2007).

Furthermore, the interaction between culture and gender provides new insights into how these factors jointly influence emotion regulation and expression. The finding that the emotional gap between Western and Eastern cultures is more pronounced among females suggests that cultural norms surrounding emotional expression may exert stronger effects on women. This aligns with research suggesting that gender roles are more strictly defined in collectivistic cultures, potentially leading to greater emotional suppression among Eastern women compared to their Western counterparts (Fischer & Manstead, 2000). Such insights extend both Hofstede's framework and emotion regulation theories by demonstrating the combined influence of cultural and gender norms on affective processes.

In addition, these findings contribute to cross-cultural neuroscience by highlighting the importance of considering both biological and sociocultural factors in emotional processing. While neuroscience has established universal mechanisms of emotion generation and regulation, this study underscores the variability introduced by cultural and gender-related socialisation processes. This reinforces the importance of interdisciplinary approaches that integrate cultural psychology, neuroscience, and emotion regulation theories to better understand human emotional diversity, especially in the textual format.

Lastly, the findings have implications for developing culturally and gender-sensitive models of emotion, which can inform future research in psychology, marketing, and artificial intelligence. By demonstrating how cultural and gender differences shape emotional experiences through multiple metrics, the study supports the need for context-aware theories of emotion regulation. These insights can refine theoretical models to better account for the dynamic interplay between individual, cultural, and neurobiological factors in shaping human emotions.

## 5.3. Practical implications

The findings of this study offer several practical contributions to the fields of consumer behaviour, marketing, and cross-cultural research. The results demonstrate that cultural differences significantly impact emotional responses, with Western consumers exhibiting higher sentiment scores, valence, arousal, and dominance than their Eastern counterparts. This suggests that marketers and advertisers should tailor their emotional appeal strategies based on cultural backgrounds. For instance, advertisements targeting Western audiences may benefit from highly arousing and dominant emotional tones, whereas those aimed at Eastern consumers might require a more subtle and balanced emotional approach.

The study also highlights the role of gender in shaping emotional experiences, as females consistently exhibit higher sentiment, valence, and arousal scores than males. This insight has important implications for advertising and product design, emphasising the need for gender-sensitive marketing strategies (Truong et al., 2020). Brands seeking to engage female audiences should focus on emotionally expressive and engaging content, while male-oriented campaigns might benefit from a more neutral or subdued emotional approach. Understanding these differences enables businesses to create more effective and resonant messages, improving customer engagement and satisfaction.

Moreover, the interaction between culture and gender further refines these insights, demonstrating that the emotional differences between Western and Eastern consumers are even more pronounced among females than males. This finding suggests that cultural differences in emotional expression are more prominent among women, making it essential for marketers to consider both cultural and gender-specific factors when designing campaigns (Truong, 2024). For example, companies launching global campaigns should customise their messaging not only by region but also by gender within each cultural group to maximise effectiveness.

These findings also have implications for customer experience management, particularly in industries where emotional engagement is crucial, such as entertainment, hospitality, and digital media. Businesses can enhance user experiences by adapting their customer interaction strategies based on the emotional tendencies of different cultural and gender groups. For instance, customer service strategies in Western markets may benefit from a more expressive and enthusiastic approach, while in Eastern markets, a more composed and harmonious interaction style may be more effective.

Additionally, the study provides valuable insights for artificial intelligence (AI) and sentiment analysis applications. AI-driven sentiment analysis tools can be refined by incorporating cultural and gender-specific emotional tendencies, leading to more accurate and context-aware sentiment predictions (Cortis, 2021). By understanding how sentiment scores, valence, arousal, and dominance vary across different demographic groups, AI developers can design better emotion recognition systems for personalised recommendations, customer service automation, and content curation.

Finally, these findings contribute to the broader discourse on cross-cultural psychology and emotional intelligence, reinforcing the idea that emotional experiences are not universal but are shaped by cultural and social contexts. Organisations involved in global business, human resources, and intercultural communication can apply these insights to foster more inclusive and effective interactions. By acknowledging and addressing the emotional variations across cultures and genders, businesses and policymakers can create strategies that enhance cross-cultural understanding, inclusivity, and engagement in diverse markets.

## 5.4. Limitations and Directions for Future Research

While this study provides valuable insights into the effects of culture and gender on customer emotional experiences, it has certain limitations that should be acknowledged. One key limitation is the reliance on text-based sentiment and emotion analysis, which may not fully capture the nuances of emotional expression. Written reviews often lack contextual cues such as tone, facial expressions, and gestures, which are crucial in understanding emotions more comprehensively (Kang, 2020). Future research could incorporate multimodal analysis by integrating text with other data sources, such as voice recordings or facial expression recognition, to gain a deeper understanding of emotional responses in digital communication, especially the video-based e-commerce platforms currently emerging.

Another limitation concerns the generalizability of the findings. The dataset used in this study consists of customer reviews, which may not represent the full spectrum of emotional experiences across different consumer interactions. Customers who leave reviews may exhibit stronger emotions, either positive or negative, compared to those who do not participate in online feedback (Yusifov & Sineva, 2022). Future research should explore more diverse data sources, including surveys, interviews, and real-time behavioural data, to validate and extend the current findings. Additionally, cross-platform comparisons could help determine whether the observed emotional patterns hold across different online environments and industries.

The study also highlights the moderating effects of culture and gender but does not extensively explore other potential influencing factors. Variables such as age, personality traits, and socioeconomic background could further shape emotional expression in consumer behaviour (Cortis, 2021). Future research could adopt a more holistic approach by incorporating these additional demographic and psychological variables into the analysis. Moreover, longitudinal studies tracking changes in emotional expression over time would provide valuable insights into evolving consumer sentiment and behaviour.

Finally, while this study employs advanced machine learning techniques for emotion detection, the accuracy and interpretability of these models remain an ongoing challenge. Different sentiment analysis models may yield varying results, and the classification of complex emotions can still be improved (Zanwar et al., 2022). Future research should refine these computational methods, possibly through more sophisticated natural language processing techniques and cross-cultural validation of emotion models (Truong, 2022). By addressing these limitations, future studies can build upon the current findings to develop a more comprehensive understanding of the role of culture and gender in shaping consumer emotional experiences.

# 6. Conclusions

Understanding consumer emotional experiences on e-commerce platforms is essential for businesses aiming to improve customer engagement and personalisation. However, three significant gaps remain in the current literature regarding the influence of gender and culture on these experiences, partly due to the recent advancements in machine learning techniques (Truong & Hoang, 2022). First, most studies have focused on basic and limited categorical emotions, while the impact of more complex emotion categories remains largely unexplored. Second, there is a lack of research on dimensional emotions, particularly the three-dimensional framework of valence, arousal, and dominance. Lastly, few studies have examined the interaction and moderating effects of gender and culture, especially in the context of text-based customer reviews, leaving an important aspect of online consumer behaviour underexplored.

This study investigates the complexity of these emotional responses by analysing how gender and cultural differences influence sentiment, valence, arousal, and dominance scores. The findings indicate that consumer emotions are more intricate than previously assumed, encompassing a broad spectrum of discrete emotions. Key differences emerge between male and female consumers, as well as between Western and Eastern cultural groups, revealing significant patterns in emotional expression across various categories.

The hypothesis testing results confirm that culture significantly affects customer emotional experiences, with Western consumers displaying higher sentiment, valence, arousal, and dominance scores compared to Eastern consumers. Additionally, gender differences are evident, with female consumers consistently exhibiting stronger emotional intensity than males across three dimensions. These variations highlight the importance of considering demographic and cultural contexts when analysing consumer feedback, as emotional responses are not uniform across different groups.

A crucial finding of this study is the interaction between gender and culture in shaping consumer emotions. The data reveal that gender-based differences in emotional expression are more pronounced in Western cultures than in Eastern ones. Specifically, the gap between male and female emotional scores is larger in Western consumer groups, a factor that has been largely overlooked in previous research. This insight underscores the necessity of examining intersectional influences rather than analysing gender and culture in isolation.

The study contributes to theoretical advancements by integrating insights from neuroscience theories, Hofstede's cultural dimension model and Emotion Regulation theory to explain these variations. It highlights how cultural norms and gender roles influence consumer emotional expression on e-commerce platforms, reinforcing the need for a multidimensional approach to understanding consumer behaviour. These findings provide a more nuanced perspective on emotional diversity, moving beyond traditional sentiment analysis to explore complex emotional categories such as admiration, amusement, and optimism. This study has suggested a framework for future research to continue exploring.

From a practical standpoint, the study offers valuable guidance for businesses aiming to refine sentiment analysis models and develop more personalised marketing strategies. By acknowledging cultural and gender-based emotional differences, companies can tailor their customer engagement approaches to better resonate with diverse audiences. These insights can enhance customer experience management, improve product recommendations, and optimise brand communication strategies, ultimately fostering stronger consumer relationships in the digital marketplace.

Plisiecki, H., & Sobieszek, A. (2024). Extrapolation of affective norms using transformer-based neural networks and its application to experimental stimuli selection. *Behavior Research Methods*, *56*(5), 4716-4731.

Posner, J., Russell, J. A., & Peterson, B. S. (2005). The circumplex model of affect: An integrative approach to affective neuroscience, cognitive development, and psychopathology. *Development and psychopathology*, *17*(3), 715-734.

Rasappan, P., Premkumar, M., Sinha, G., & Chandrasekaran, K. (2024). Transforming sentiment analysis for e-commerce product reviews: Hybrid deep learning model with an innovative term weighting and feature selection. *Information Processing & Management*, *61*(3), 103654.

Russell, J. A. (2003). Core affect and the psychological construction of emotion. *Psychological review*, *110*(1), 145.

Safdar, S., Friedlmeier, W., Matsumoto, D., Yoo, S. H., Kwantes, C. T., Kakai, H., & Shigemasu, E. (2009). Variations of emotional display rules within and across cultures: A comparison between Canada, USA, and Japan. *Canadian Journal of Behavioural Science/Revue canadienne des sciences du comportement*, *41*(1), 1.

Saunders, M. N. K. (2015). Research Methods for Business Students. In P. Lewis & A. Thornhill (Eds.), (7th ed ed.): Harlow, United Kingdom : Pearson Education Limited.

Scherer, K. R., Banse, R., & Wallbott, H. G. (2001). Emotion inferences from vocal expression correlate across languages and cultures. *Journal of Cross-Cultural Psychology*, *32*(1), 76-92.

Simon, R. W., & Nath, L. E. (2004). Gender and emotion in the United States: Do men and women differ in self-reports of feelings and expressive behavior? *American journal of sociology*, *109*(5), 1137-1176.

Susanto, Y., Livingstone, A. G., Ng, B. C., & Cambria, E. (2020). The hourglass model revisited. *IEEE Intelligent Systems*, *35*(5), 96-102.

Thelwall, M. (2018). Gender bias in sentiment analysis. *Online Information Review*, *42*(1), 45-57.

Truong, V. (2022). Natural language processing in advertising–a systematic literature review. 2022 5th Asia Conference on Machine Learning and Computing (ACMLC),

Truong, V. (2023). Optimizing mobile in-app advertising effectiveness using app publishers-controlled factors. *Journal of Marketing Analytics*, 1-19.

Truong, V. (2024). Textual emotion detection–A systematic literature review.

Truong, V., & Hoang, V. (2022). Machine learning optimization in computational advertising—A systematic literature review. *Intelligent Systems Modeling and Simulation II: Machine Learning, Neural Networks, Efficient Numerical Algorithm and Statistical Methods*, 97-111.

Truong, V., Nkhoma, M., & Pansuwong, W. (2020). Enhancing the effectiveness of mobile in-app programmatic advertising using publishers-controlled factors. Proceedings of the 2020 the 3rd International Conference on Computers in Management and Business,

Tsai, J. L., & Clobert, M. (2019). Cultural influences on emotion: Established patterns and emerging trends.

Vonk, A., & Silva, V. F. (2024). The Hofstede Model: Understanding a Multicultural Environment. In *Cultural Confluence in Organizational Change: A Portuguese Venture in Angola* (pp. 47-76). Springer.

Wang, B., Shen, C., Cai, Y., Dai, L., Gai, S., & Liu, D. (2023). Consumer culture in traditional food market: the influence of Chinese consumers to the cultural construction of Chinese barbecue. *Food Control*, *143*, 109311.

Wang, X., Guo, J., Wu, Y., & Liu, N. (2020). Emotion as signal of product quality: Its effect on purchase decision based on online customer reviews. *Internet Research*, *30*(2), 463-485.
38